\documentclass[a4paper,10pt]{article}

\usepackage{jcappub} % for details on the use of the package, please see the JINST-author-manual

\usepackage{amsmath,amssymb}
\usepackage{graphicx}
\usepackage{dcolumn}
\usepackage{bm,color}
\usepackage{hyperref}
\usepackage{accents}
\usepackage{amssymb,float}
\usepackage{amsmath}
\usepackage{multirow}
\usepackage{siunitx}
\usepackage{tabularx}
\usepackage{booktabs}
\usepackage{url}
\usepackage{bm}
\usepackage{braket}
\usepackage{tikz}
\usetikzlibrary{positioning}
\def\layersep{2.5cm}
\usepackage{url}
\usepackage{accents}
\usepackage[sort&compress]{natbib}
\usepackage{subfigure}
\usepackage{soul}

\graphicspath{ {figs/} }
\hypersetup{colorlinks=true,citecolor=blue, linkcolor=red, urlcolor=blue}%, hidelinks}

\title{\boldmath A possible late-time transition of $M_B$ inferred via neural networks}

\author[a]{Purba Mukherjee}
\author[b]{Konstantinos F. Dialektopoulos}
\author[c,d]{Jackson Levi Said}
\author[c,d]{Jurgen Mifsud}

\affiliation[a]{Centre for Theoretical Physics, Jamia Millia Islamia, New Delhi - 110025, India}
\affiliation[b]{Physics and Applied Mathematics Unit, Indian Statistical Institute, Kolkata - 700108, India}
\affiliation[c]{Department of Mathematics and Computer Science, Transilvania University of Brasov, Eroilor 29, Brasov, Romania}
\affiliation[d]{Institute of Space Sciences and Astronomy, University of Malta, Msida, Malta}
\affiliation[e]{Department of Physics, University of Malta, Msida, Malta}

\emailAdd{purba16@gmail.com}
\emailAdd{kdialekt@gmail.com}
\emailAdd{jackson.said@um.edu.mt}
\emailAdd{jurgen.mifsud@um.edu.mt}

\abstract{The strengthening of tensions in the cosmological parameters has led to reconsidering the fundamental aspects of standard cosmology. The tension in the Hubble constant can also be viewed as a tension between local and early Universe constraints on the absolute magnitude $M_B$ of Type Ia supernova. In this work, we reconsider the possibility of a variation of this parameter in a model-independent way. We employ neural networks to agnostically constrain the value of the absolute magnitude as well as assess the impact and statistical significance of a variation in $M_B$ with redshift from the Pantheon+ compilation, together with a thorough analysis of the neural network architecture. We find an indication for a possible transition redshift at the $z\approx 1$ region.
}

\arxivnumber{2402.10502}

\begin{document}

\maketitle

\section{Introduction} \label{sec:intro}

The $\Lambda$CDM concordance model has provided a good description of both astrophysical and cosmological phenomenology for several decades \cite{Peebles:2002gy,Copeland:2006wr}. In this regime, cold dark matter (CDM) acts to stabilize galaxies \cite{Baudis:2016qwx,XENON:2018voc} while the late time accelerated expansion of the Universe \cite{Riess:1998cb,Perlmutter:1998np} is driven by a cosmological constant \cite{Mukhanov:991646}. The standard model of cosmology is completed by the inclusion of an inflationary epoch in the very early Universe \cite{Guth:1980zm,Linde:1981mu}. On the other hand, the concordance model has several underlying problems that includes the theoretical value of the cosmological constant \cite{Weinberg:1988cp}, the UV completeness of the theory \cite{Addazi:2021xuf}, and questions about the direct measurement of CDM particles \cite{Gaitskell:2004gd,LUX:2016ggv}. Recently, an observation-driven inconsistency has arisen in which the predictions of the $\Lambda$CDM model for the value of the Hubble constant (based on early Universe measurements) and the value obtained through surveys appear to diverge \cite{Abdalla:2022yfr}. This Hubble tension is part of a larger cosmological tensions problem in which different cosmological parameters have differing predicted and observed values \cite{DiValentino:2020vhf,DiValentino:2020zio,DiValentino:2020vvd,Staicova:2021ajb,DiValentino:2021izs,Perivolaropoulos:2021jda,Krishnan:2021jmh,SajjadAthar:2021prg,CANTATA:2021ktz,DiValentino:2022oon,Colgain:2022tql,Anderson:2023aga}.

There has been a dramatic increase in the reporting of the divergence of Hubble constant values in recent years. The discrepancy between direct and indirect measurements of $H_0$ has reached a critical threshold in statistical significance \cite{Aghanim:2018eyx,Schoneberg:2022ggi,ACT:2023kun}. For the early Universe, the latest reported Hubble constant by the Planck and ACT collaborations are respectively $H_0^{\rm P18} = 67.4 \pm 0.5 \,{\rm km\, s}^{-1} {\rm Mpc}^{-1}$ \cite{Aghanim:2018eyx} and $H_0^{\rm ACT-DR4} = 67.9 \pm 1.5 \,{\rm km\, s}^{-1} {\rm Mpc}^{-1}$ \cite{ACT:2020gnv}. While in the late Universe, direct measurements have produced corresponding values by the SH0ES Team and the Carnegie Supernova Project gives respective reported values of $H_0^{\rm S22} = 73.5 \pm 1.1 \,{\rm km\, s}^{-1} {\rm Mpc}^{-1}$ \cite{Brout:2022vxf} and $H_0^{\rm CSP23} = 73.22 \pm 1.45 \,{\rm km\, s}^{-1} {\rm Mpc}^{-1}$ \cite{Uddin:2023iob}. Other reported values based on strong lensing by TDCOSMO gives a Hubble constant in the range $H_0^{\rm TDCOSMO23} = 73.3 \pm 5.8 \,{\rm km\, s}^{-1} {\rm Mpc}^{-1}$ \cite{Shajib:2023uig}. Each survey shows a high level of internal consistency, the tension arises between survey results, and is particularly poignant for surveys that rely on the $\Lambda$CDM concordance model in making their estimates \cite{Bernal:2016gxb}.

There has been a diverse range of reactions to the cosmological tensions problem with the prospect of systematics being exceedingly unlikely since the problem has appeared across a wide range of unconnected surveys. To tackle this growing issue, there have been a number of proposals of modifications to the concordance model in which the early Universe \cite{Poulin:2023lkg}, the neutrino sector \cite{DiValentino:2021imh}, or the underlying gravitational model \cite{Cai:2019bdh,Addazi:2021xuf,CANTATA:2021ktz,LeviSaid:2021yat} have been changed. Nonetheless, the absence of a universally accepted cosmological model makes linking physical theory to current observational data challenging. To address this, researchers construct viable cosmological models directly from observational data, reversing the traditional method of deriving evolution from Einstein's field equations. This reverse-engineering approach \textit{aka} reconstruction utilizes observational data to directly infer the distribution of matter sector, a technique widely used in modern cosmology. Key quantities often studied include the dark energy equation of state parameter \( w_d \) \cite{Saini:1999ba,Sahni:2006pa} and the quintessence potential \( V(\phi) \) \cite{Huterer:1998qv, Starobinsky:1998fr, Huterer:2000mj}. The direct approach is a parametric reconstruction where the cosmological quantities are represented as simple functions of redshift, along with some model parameters that are estimated using observational data \cite{Huterer:2000mj, Chevallier:2000qy, Linder:2002et, Gerke:2002sx, Gong:2006gs, Scherrer:2007pu, Scherrer:2008be}. However, this method can introduce bias due to its reliance on a specific functional form for the redshift dependence.

Another approach that is gaining increased interest in the literature is that of using nonparametric approaches to reconstruct the evolution of cosmological parameters using methods that are independent of physical models. Currently, the most studied implementation of these approaches has taken the form of Gaussian processes \cite{10.5555/1162254} which is based on the use of a kernel function that characterizes the covariance relationships between the observed data. The kernels are defined on nonphysical hyperparameters which can then be fit using traditional techniques. There have been a plethora of works in this direction, which includes the reconstruction of various classes of cosmological models \cite{Holsclaw:2010nb,Holsclaw:2010sk,Holsclaw:2011wi,2012JCAP...06..036S,Shafieloo:2012ht,Seikel:2013fda,Busti:2014aoa,Busti:2014dua,Cai:2019bdh,Briffa:2020qli,Mukherjee:2021ggf,Mukherjee:2021epjc,Bernardo:2021mfs,Bernardo:2021cxi,LeviSaid:2021yat,Bernardo:2021qhu,Benisty:2020kdt,Escamilla-Rivera:2021rbe,Mukherjee:2020vkx,Benisty:2022psx,Bernardo:2022pyz,Hwang:2022hla,Ren:2022aeo,Shah:2023rqb,Mukherjee:2023lqr,Banerjee:2023rvg,Dinda:2024ktd,Mukherjee:2024ryz}. These approaches are interesting, but as an algorithm, GPs possess several shortfalls such as overfitting issue at low redshifts, as well as an over-reliance on the form of the kernel function. As an alternative to GPs, Genetic Algorithms (GAs) have also been widely utilized for non-parametric reconstruction of cosmological functions using analytical forms as priors \cite{Bogdanos:2009ib,Nesseris:2010ep,Nesseris:2012tt,Arjona:2019fwb,Arjona:2020kco,Arjona:2021hmg}. Besides, methods like the iterative smoothing or local regression  \cite{Shafieloo:2005nd, Shafieloo:2007cs, LHuillier:2016mtc, Shafieloo:2018gin, Koo:2021suo}, weighted polynomial regression \cite{Gomez-Valent:2018hwc, Gomez-Valent:2018gvm}, principal component analysis \cite{Crittenden:2005wj,Crittenden:2011aa,Wang:2015wga,Sharma:2020unh}, nodal and free-form reconstruction methods \cite{AlbertoVazquez:2012ofj,Vazquez:2012ux,Gerardi:2019obr,Hazra:2019wdn,Krishak:2021fxp,Escamilla:2023shf,Mukherjee:2024cfq} have gained traction over the recent years. Although each of these methods has its limitations, particularly the interpretation of uncertainties can be more complex compared to GPs. 

Recently, machine learning approaches that have shown promise in confronting the problem of removing model reliance in cosmological data are artificial neural networks (ANN) \cite{10.2307/j.ctt4cgbdj,Escamilla-Rivera:2019hqt} where the number of nonphysical hyperparameters is dramatically increased. Besides, there is no appearance of the kernel covariance function compared to GPs. In this method, neurons are modeled on their biological analogs, which are then organized into layers that pass signals or inputs through the entirety of the network. This could directly take the form of redshift inputs and Hubble parameter outputs \cite{aggarwal2018neural,Wang:2020sxl,Gomez-Vargas:2021zyl}. One example of an ANN implementation in the cosmological context is Ref.~\cite{Wang:2019vxv} where the Hubble diagram was reconstructed for a number of data sets, this was then further developed in Ref.~\cite{Dialektopoulos:2021wde} to include more complexity in the training data as well as null test analysis of the reconstructions. In Ref.~\cite{Dialektopoulos:2023dhb} this was further solidified with the native inclusion of the underlying covariances in cosmological data forming part of the training process in which the Hubble diagram is produced. ANNs have also been used to devise an approach in which the derivatives of reconstructed parameters, and their associated uncertainties, can be calculated \cite{Mukherjee:2022yyq} which has led the way to making these methods competitive with Gaussian processes in that they can be used to reconstruct classes of modified cosmological models \cite{Dialektopoulos:2023jam}. Recently, LADDER (Learning Algorithm for Deep Distance Estimation and Reconstruction) \cite{Shah:2024slr} has utilized deep neural networks (DNN) to reconstruct the cosmic distance ladder, highlighting the potential benefits of advanced learning techniques in enhancing optimal information extraction from cosmological data.

In this work, we use the robustness of reconstructions based on ANN architectures to probe the potential evolution of the absolute magnitude of Type Ia supernovae events (SNIa). It has been suggested in the literature that the Hubble tension may be interpreted as a tension in the absolute magnitude \cite{Camarena:2021jlr,Camarena:2022iae}. Indeed, one can certainly transform the Hubble tension into a tension in this value; however, this may imply new physics at the level of the behaviour of astrophysical events. A consequence of this tension in the absolute magnitude would be that one cannot use priors in their analysis of modified cosmological models \cite{Efstathiou:2021ocp}. Another consequence of this perspective of the tension in the expansion rate of the Universe at present is that one may devise proposals of models through which the absolute magnitude may vary in redshift space. Through this viewpoint, the instrument of nonparametric reconstruction techniques can be used to probe the possible evolution of the absolute magnitude in its own right. Model-independent constraints on the absolute magnitude was attempted by Dinda \& Banerjee \cite{Dinda:2022jih} employing Gaussian processes (GP), where an extensive analysis was carried out with the choice of different kernels and mean functions. As a rigorous extension of \cite{Dinda:2022jih}, joint constraints on some cosmological nuisance parameters were attempted in Refs. \cite{Dinda:2022vmb, Dinda:2023svr} employing different combinations of data sets. Similar efforts in these directions were carried out in Ref. \cite{Mukherjee:2021epjc,Mukherjee:2021kcu,Banerjee:2023evd}. Recent attempts to explore the inherent degeneracy between the parameters, $M_B$ and $r_d$ (comoving sound horizon at drag epoch) with deep neural networks, was carried out in Ref. \cite{Shah:2024slr}. Indeed, there have been several physically motivated proposals for the potential evolution of the absolute magnitude parameter coming from phenomena. In Ref.~\cite{Benisty:2022psx}, the constancy of the absolute magnitude was probed comparatively through the reconstruction of the absolute magnitude using three general methods composed of ANNs, two kernel choices of GP reconstruction, and four models taken from the literature \cite{2009A&A...506.1095L,Tutusaus:2017ibk}. Similar efforts to constraining $M_B$ or revisiting the constancy of $M_B$ in a non-flat universe were carried out by \cite{Mukherjee:2020vkx, Favale:2023lnp} for different combinations of datasets. However, results showed no significant evidence to merit an evolution in this parameter. It deserves mention that some of these non-parametric studies employed a combined approach in which baryon acoustic oscillation (BAO) data was used to isolate the luminosity distance and thus infer a value of the absolute magnitude. However, the BAO data is not completely model-independent and may have inserted a preference for this conclusion. 

In the current work, we are motivated to reconsider the uncalibrated latest Type Ia supernovae data, namely the Pantheon+ compilation, and use this to reconstruct the apparent magnitudes and their derivatives in a cosmology-independent way using neural networks. Through this route, we can constrain the value of the absolute magnitude $M_B$ without making any assumptions on the underlying cosmological model whatsoever. This is achieved by comparing the cosmic chronometer (CC) measurements of the Hubble parameter with the ANN reconstructed Hubble diagram from Pantheon+, thereby computing the absolute magnitude value in an agnostic way. We do this by introducing the technical details of the ANN approach in Sec.~\ref{sec:ann_intro}, which directly leads to the reconstruction methodology adopted as detailed in Sec.~\ref{sec:methodology}. In Sec.~\ref{sec:results}, we show the cosmology-independent constraints on $M_B$ that ANNs can achieve using this approach. Finally, we give a summary of our main conclusions and future work in Sec.~\ref{sec:conclusion}.

\section{Artificial Neural Networks}\label{sec:ann_intro} 

This section briefly describes the method by which ANN architectures \cite{2015arXiv151107289C} have been adopted. Inspired from biological neural networks, ANNs are built as a collection of neurons that are organized into layers. An input layer is usually used to enter the data into the network, an output layer is used to extract the parameters we want to learn their behaviour, while there is a series of consecutive layers of neurons in between, usually called \textit{hidden layers}, that depend on a number of hyperparameters and are being optimized to best mimic the real data processes.

In our case, the input layer will accept redshift values while the output layer gives the supernovae apparent magnitude $m(z)$ for that redshift and its uncertainty. In this way, an input signal, or redshift value, traverses the whole network to train the network, i.e. optimize the hyperparameters of each neuron and then produce these outputs. To illustrate this architecture, we show a two hidden layer ANN for a generic cosmological parameter $\Upsilon(z)$ (which in our case is $m(z)$) and its corresponding uncertainty $\sigma_\Upsilon^{}(z)$ (or $\sigma _m$ in our case) in Fig. \ref{fig:ANN_structure}, where the neurons are denoted by $\mathfrak{n}_k$ and $\mathfrak{m}_k$. Here, the ANN is structured so that a linear transformation (composed of linear weights and biases) is applied for each of the different layers.

\tikzset{%
  every neuron/.style={
    circle,
    fill=green!70,
    minimum size=32pt, inner sep=0pt
  },
  mid neuron/.style={
    circle,
    fill=blue!40,
    minimum size=32pt, inner sep=0pt
  },
  last neuron/.style={
    circle,
    fill=red!60,
    minimum size=32pt, inner sep=0pt
  },
  neuron missing/.style={
    draw=none,
    fill=none,
    scale=4,
    text height=0.333cm,
    execute at begin node=\color{black}$\vdots$
  },
}
\begin{figure}[t!]
    \centering
    \begin{tikzpicture}[shorten >=1pt,->,draw=black!50, node distance=\layersep]
    \tikzstyle{annot} = [text width=5em, text centered]

\foreach \m/\l [count=\y] in {1}
  \node [every neuron/.try, neuron \m/.try] (input-\m) at (0,-1.5*\y) {};

\foreach \m [count=\y] in {1,2,3,missing,4}
  \node [mid neuron/.try, neuron \m/.try ] (hidden1-\m) at (3.5,2-\y*1.5) {};

\foreach \m [count=\y] in {1,2,3,missing,4}
  \node [mid neuron/.try, neuron \m/.try ] (hidden2-\m) at (6.5,2-\y*1.5) {};

\foreach \m [count=\y] in {1,2}
  \node [last neuron/.try, neuron \m/.try ] (output-\m) at (10,1.25-2*\y) {};

\foreach \name / \y in {1}
    \path[yshift=-.1cm] node[above] (input+\name) at (0,-1.6\name) {\large$z$};

\foreach \l [count=\i] in {1,2,3,k}
  \node at (hidden1-\i) {\large$\mathfrak{n}_\l$};
  
\foreach \l [count=\i] in {1,2,3,k}
  \node at (hidden2-\i) {\large$\mathfrak{m}_\l$};

\foreach \name / \y in {{\large $\Upsilon(z)$} / 1, {\large$\sigma_\Upsilon^{}(z)$} / 2}
        \path[yshift=-.1cm] node[above, right of=H-3] 
        (output-\y) at (7.5,1.35-2*\y) {\name};

\foreach \i in {1}
  \foreach \j in {1,2,3,...,4}
    \draw [->] (input-\i) -- (hidden1-\j);

\foreach \i in {1,2,3,...,4}
  \foreach \j in {1,2,3,...,4}
    \draw [->] (hidden1-\i) -- (hidden2-\j);

\foreach \i in {1,2,3,...,4}
  \foreach \j in {1,2}
    \draw [->] (hidden2-\i) -- (output-\j);
    
\node (1) at (0,3){\large Input};
\node (1) at (0,2.55){\large layer};

\node (1) at (5,3){\large Hidden};
\node (1) at (5,2.55){\large layers};

\node (1) at (10,3){\large Output};
\node (1) at (10,2.55){\large layer};
%\foreach \l [count=\x from 0] in {\large Input, \large Hidden, \large Output}
%    \node [align=center, above] at (\x*5,2) {\l \\ \large layer};

\end{tikzpicture}
    \caption{The adopted ANN architecture is shown, where the input is the redshift of a cosmological parameter $\Upsilon(z)$, and the outputs are the corresponding value and error of $\Upsilon(z)$.}
\label{fig:ANN_structure}
\end{figure}
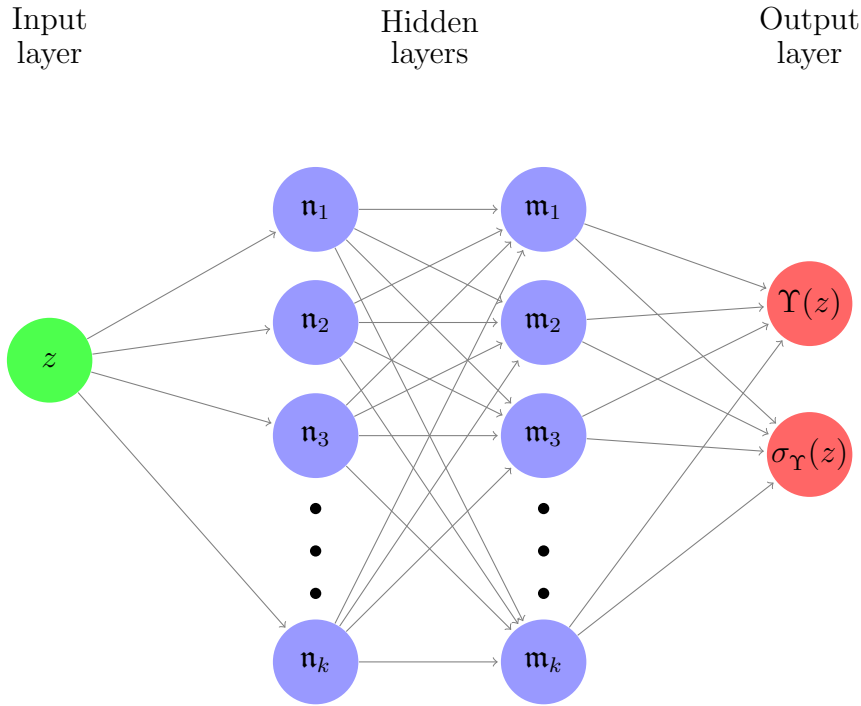

Neurons are triggered by an activation function that, across the larger number of neurons, can be used to model the complex relationships that feature in the data. In our work, we used the Exponential Linear Unit (ELU) \cite{2015arXiv151107289C} as the activation function, specified by
\begin{equation}
    f(x) = 
  \begin{cases} 
   {x} & \text{if } x>0 \\
   {\alpha(e^x-1)} & \text{if } x \leq 0
  \end{cases}\,,
\end{equation}
where $\alpha$ is a positive hyperparameter that controls the value to which an ELU saturates for negative net inputs, which we set to unity. Besides being continuous and differentiable, this function does not act on positive inputs while negative inputs tend to be closer to unity for more negative input values. The choice of the activation function for the hidden layers usually has to do with the type of the network, i.e. multilayer perceptron, convolutional, recurrent, etc, while the choice for the output layer activation function has to do with the type of the problem (classification or regression); more details can be found in \cite{DeepLearning-book}.

The linear transformations and activation function produce a huge number of so-called hyperparameters which are nonphysical and can be optimized using training data so that the larger system mimics some physical process as closely as possible. Indeed, the full number of hyperparameters is even larger since there are additional hyperparameters related to the connection of the neurons which drastically increases the freedom of the ANN systems. The process by which training takes place involves these hyperparameters being optimized by comparing the predicted result $\hat{\Upsilon}$ with the real values $\Upsilon$ (training data) so that their difference is minimized. This minimization is called the loss function. This function is minimized by fitting methods, such as gradient descent, which fixes the hyperparameters for specific data sets, or a combination of them. In our work, we adopt Adam's algorithm \cite{2014arXiv1412.6980K} as our optimizer, which is a modification of the gradient descent method that has been shown to accelerate convergence. 

The direct absolute difference between the predicted ($\hat{\Upsilon}$) and training ($\Upsilon$) outputs summed for every redshift is called the L1 loss function. This is akin to the log-likelihood function for uncorrelated data in a Markov Chain Monte Carlo (MCMC) analysis and is a very popular choice for ANN architectures. There are other choices such as the mean squared error (MSE) loss function which minimizes the square difference between $\hat{\Upsilon}$ and $\Upsilon$, and the smooth L1 (SL1) loss function which uses a squared term if the absolute error falls below unity and absolute term otherwise; one can see a comparison of different loss functions in \cite{Dialektopoulos:2021wde}. Thus, it is at the level of the loss function that complexities in the data are inserted into the ANN through optimization in the training of the numerous hyperparameters that make up the system. In our work, we do this by assuming a loss function that is more akin to correlated data when considering log-likelihood functions for MCMC analyses. To that end, we incorporate the covariance matrix $\mathcal{C}$ of a data set by taking a loss function, given as
\begin{equation}
    {\rm L_{\chi^2}} = \sum_{i,j} \left[m_{\rm obs}(z_i) - m_{\rm pred}(z_i)\right]^\text{T} \mathcal{C}_{ij}^{-1} \left[m_{\rm obs}(z_j) - m_{\rm pred}(z_j)\right] \,, \label{eq:chi2_loss}
\end{equation}
where $\mathcal{C}_{ij}$ is the total noise covariance matrix of the data, which includes the statistical noise and systematics.

ANNs that feature at least one hidden layer can approximate any continuous function for a finite number of neurons, provided the activation function is continuous and differentiable \cite{HORNIK1990551}, which means that ANNs are applicable to the setting of cosmological data sets. In this work, we utilize  \texttt{PyTorch}\footnote{\url{https://pytorch.org/docs/master/index.html}} for our ANN implementation. The code was run on GPUs to speed up the computational time, as well as making use of batch normalization \cite{2015arXiv150203167I} prior to every layer which further accelerates the convergence.

\section{Reconstruction Methodology} \label{sec:methodology}

In this section, we discuss the data sets used in this analysis together with our training and validation strategy for structuring the ANN system. Finally, we lay out the physical framework on which we perform this model-independent constraint analysis on the evolution of $M_B$.

\subsection{Observational Data sets \label{subsec:datasets}}

As mentioned in the introduction, in this work, we consider two sets of observational data. We consider the latest Pantheon+ compilation for SNIa observations \cite{Brout:2021mpj, Riess:2021jrx, Scolnic:2021amr} from 1701 light curves that represent 1550 distinct SNIa spanning the redshift range $z<2.3$. This data set comprises apparent magnitude measurements $m$ with their associated statistical uncertainties, tabulated at different redshifts. It also includes a covariance matrix detailing the systematic errors or correlations in the measurements. Recent investigations with Pantheon+ have highlighted concerns, including a possible 5\% overestimation of uncertainties in the covariance matrix \cite{Keeley:2022iba} and the potential presence of an unaddressed systematic effect—volumetric redshift scatter—that could suggest a physical shift in the absolute magnitude of SNIa at redshift $z = 0.005$ \cite{Perivolaropoulos:2023iqj}.

We also take into account the latest data set of 32 CC $H(z)$ measurements \cite{Stern:2009ep,Moresco:2012jh, Moresco:2016mzx, Borghi:2021rft, Ratsimbazafy:2017vga, Moresco:2015cya, Zhang:2012mp}, along with the full covariance matrix that includes systematic and calibration errors, as reported in Ref. \cite{Moresco:2020fbm}. These data do not assume any particular cosmological model, but depend on the differential ages technique between galaxies, covering the redshift range up to $z \sim 2$. The CC method, although a new emerging cosmological probe that provides model-independent estimates of the Universe's expansion history, its reliability is often debated within the cosmology community \cite{Abdalla:2022yfr}. Major large-scale cosmological surveys like the SDSS, DESI, or DES, prioritize other observational techniques like SNIa, BAO, and galaxy clustering over the CC method. Its reliance on the age of the Universe, being an integrated quantity of $H(z)$, faces limitations in statistical power and susceptibility to systematic errors, such as the calibration of Stellar Population Synthesis (SPS) models and the accurate selection of CC samples to minimize contamination \cite{Moresco:2023zys}. Although we incorporate this data into our analysis, its precision and accuracy are compromised by uncertainties in metallicity, biases from star formation history, and systematic errors from spectral models \cite{Moresco:2022phi}. These factors introduce potential biases and interpretational challenges (see Ref. \cite{Gomez-Valent:2018hwc, Haridasu:2018gqm, Mukherjee:2020vkx} for the possible impact of CC systematics on cosmological results), reducing its robustness and ability to independently drive our conclusions. Therefore, our findings are based on the assumption of this data's reliability, recognizing the need for further validation and scrutiny.

\subsection{ANN Training and Validation \label{subsec:training}}

Following the prescription outlined in Sec.~\ref{sec:ann_intro}, we train our neural network using the Pantheon+ SNIa dataset \cite{Brout:2021mpj, Riess:2021jrx, Scolnic:2021amr}. To optimize the network's performance, we split the dataset into training (70\%) and validation (30\%) sets. To incorporate the covariance matrix of the Pantheon+ dataset into the training algorithm, the $\chi^2$ loss function, defined in Eq.~\eqref{eq:chi2_loss} is minimized. During training, the number of iterations is set to 30,000 at every epoch where a batch size of 32 was considered. The covariance sub-matrices equivalent for every training batch were carefully chosen from the actual data ensuring that these sub-matrices are positive semi-definite. 

\begin{figure}
\centering
\includegraphics[width=0.45\textwidth]{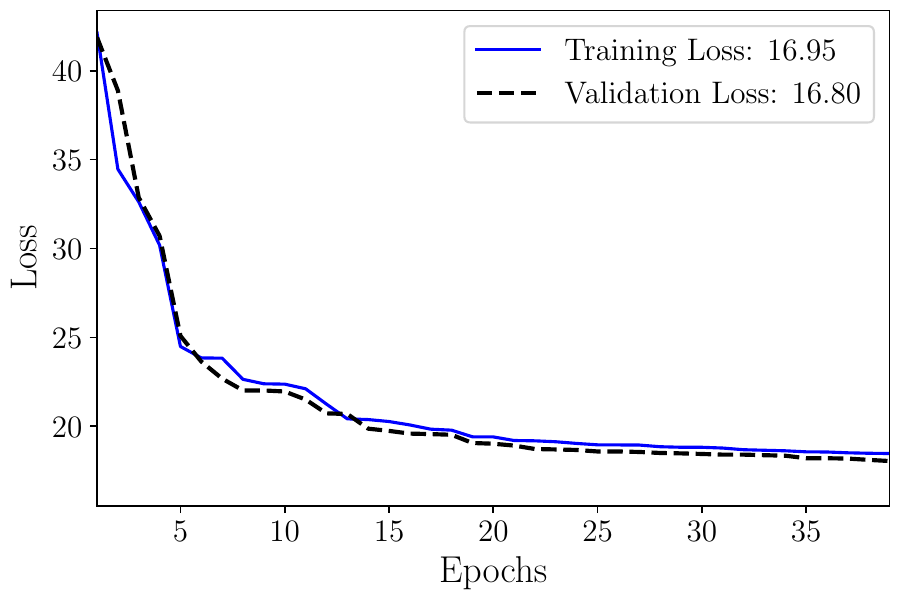} 
\includegraphics[width=0.45\textwidth]{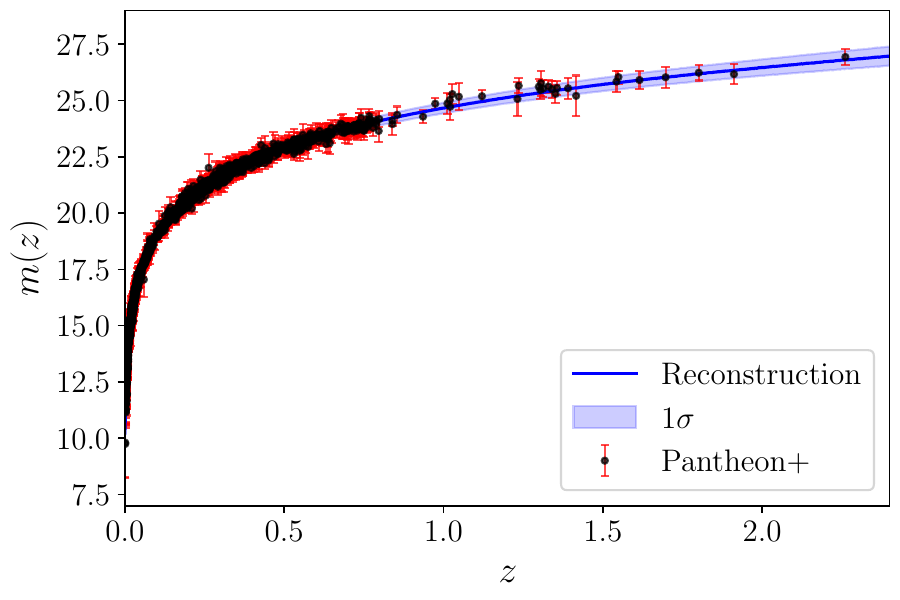}
\caption{Plots illustrating the evolution of training loss and validation loss across different epochs (left panel). ANN reconstruction of the Pantheon+ SNIa apparent magnitudes $m(z)$ as a function of the redshift $(z)$.} \label{fig:ANN_plots}
\end{figure}

The left panel of Fig. \ref{fig:ANN_plots} showcases the respective training and validation losses across different epochs. To identify the best network architecture, we employ an early stopping criterion based on the ongoing comparison with the validation loss. This strategy helps us prevent overfitting and ensures the selection of a model that generalizes well to unseen data. We find that the optimal network configuration features two hidden layers with 128 neurons each. 

Our comprehensive approach to training the neural network ensures robust performance in predicting SNIa apparent magnitudes across varying redshifts. To demonstrate the effectiveness of our algorithm, we present the reconstructed mean and 1$\sigma$ uncertainties as functions of redshift $z$ in the right panel of Fig. \ref{fig:ANN_plots}. We employ a Monte Carlo dropout approach \cite{Dialektopoulos:2023dhb}, which involves running multiple forward passes with dropout during inference, generating a distribution of predictions that accounts for the inherent uncertainty in the model. This method ensures accurate reconstructions and quantifies prediction uncertainties of $m(z)$ at any arbitrary redshifts, making our algorithm a valuable tool for robust cosmological analyses.

\begin{figure}
\centering
\includegraphics[width=0.475\textwidth]{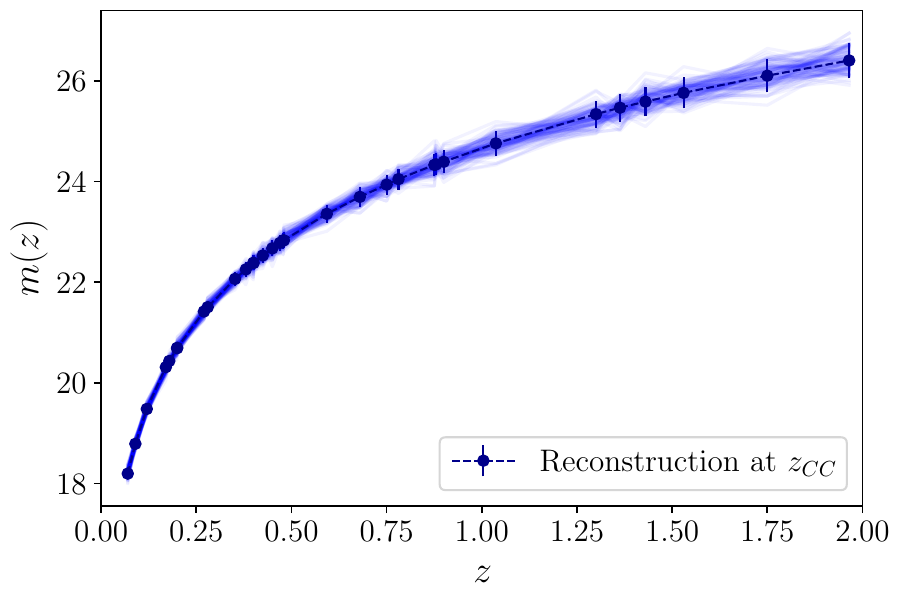}
\includegraphics[width=0.475\textwidth]{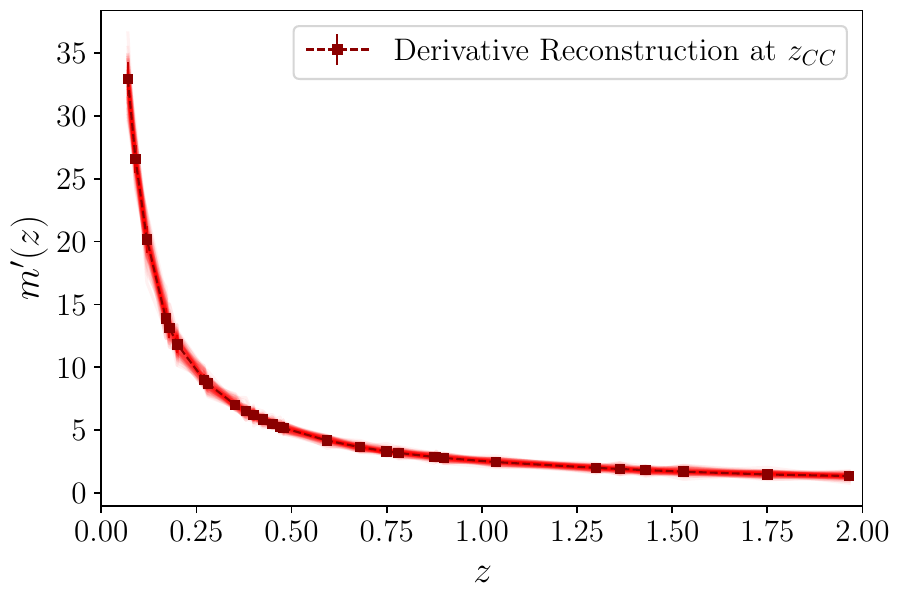}
\caption{ANN reconstruction of the Pantheon+ apparent magnitudes $m(z)$ (left panel) and its corresponding derivatives $m^{\prime}(z)$ [right panel] at the CC redshifts $(z_{\text{CC}})$.} \label{fig:mz_cc_recon}
\end{figure} 

\begin{figure}
\centering
\includegraphics[width=\textwidth]{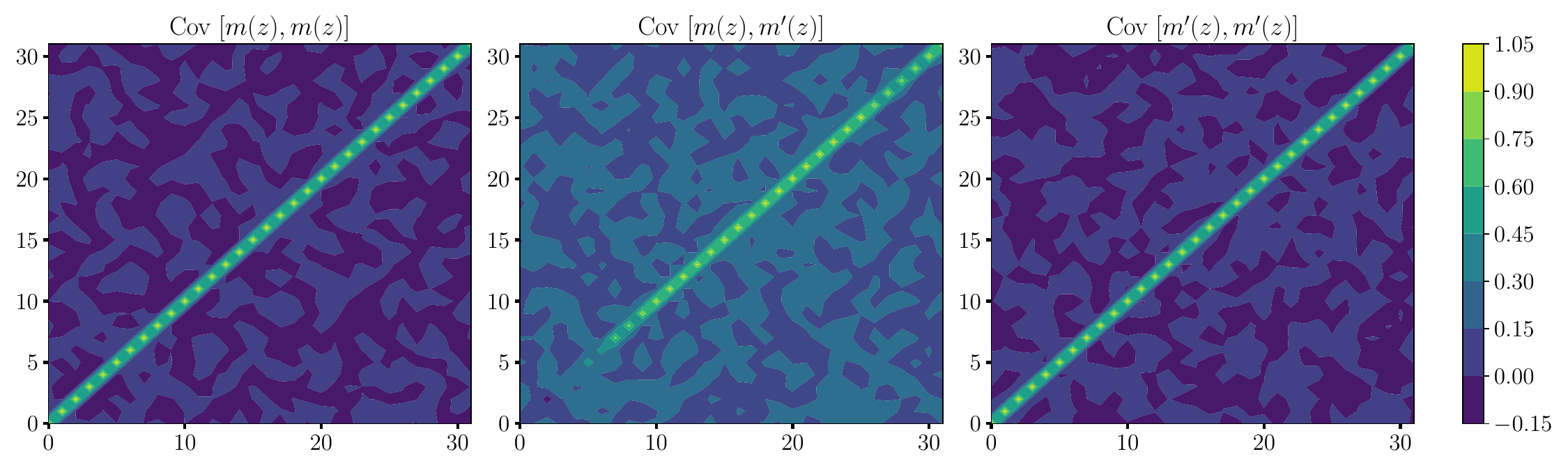}
\caption{Normalized covariance matrices between ANN reconstruction of the Pantheon+ SN-Ia apparent magnitudes $m(z)$ and its corresponding derivatives $m^{\prime}(z)$ at the CC redshifts $(z_{\text{CC}})$.} \label{fig:mz_cc_cov_recon}
\end{figure}

\subsection{Theoretical Framework \label{subsec:theory}}

In a spatially flat Friedmann-Lema\^{i}tre-Robertson-Walker universe, the luminosity distance is related to the Hubble parameter $H(z)$ at some redshift $z$, as,
\begin{equation} \label{eq:define_dL}
    d_L(z) = c (1+z) \int_0^z \frac{d \tilde{z}}{H(\tilde{z})}.
\end{equation}
The observed luminosity of SNIa, from a specific redshift, is related to the apparent peak magnitude $m$ via the following relation, independent of any physical model as,
\begin{equation} \label{eq:m_sn}
    m(z) - M_B = 5 \log_{10} \left[ \frac{d_L(z)}{\text{1 Mpc}}\right] + 25 \, .
\end{equation} 
From Eq. \eqref{eq:m_sn}, we can rewrite the luminosity distance as,
\begin{equation} \label{eq:dL_from_sn}
d_L(z) = 10^{\frac{1}{5} \left[ m(z) - M_B - 25 \right]}.
\end{equation}
Furthermore, we can compute $d_L^\prime$, the first order derivative of $d_L$ with respect to the redshift $z$ as,
\begin{equation} \label{eq:dLprime_from_sn}
 d_L^\prime(z) =  \frac{\log(10)}{5} 10^{-\frac{M_B}{5}} m^\prime(z).    
\end{equation}
Combining Eq. \eqref{eq:dL_from_sn} and Eq. \eqref{eq:dLprime_from_sn} with Eq. \eqref{eq:define_dL}, we can express the Hubble parameter as,
\begin{equation} \label{eq:Hubble_from_sn}
    H(z) = \frac{c (1+z)^2}{(1+z)d_L^\prime(z) -  d_L(z)} .
\end{equation}
In this way, we can derive the Hubble parameter $H(z)$, from the Pantheon+ apparent magnitudes $m$ and its corresponding derivatives $m^{\prime}$ employing specific values of $M_B$. A similar approach is used in the following section in order to get constraints on $M_B$ in a cosmological model-independent way.

\section{Results and Discussions \label{sec:results}}

The ANN architecture is configured in the preceding section with a focus on producing a robust architecture that can reconstruct the possible evolution of $M_B$. Here we perform this analysis by using the evolution profiles that the ANN produces. Firstly, our interest is in determining a constant value for the absolute luminosity, and secondly, we explore the possibility of a free evolution through which it turns out that a mild preference for a transition point is determined.

\begin{figure}
\centering
\includegraphics[width=0.2\textwidth]{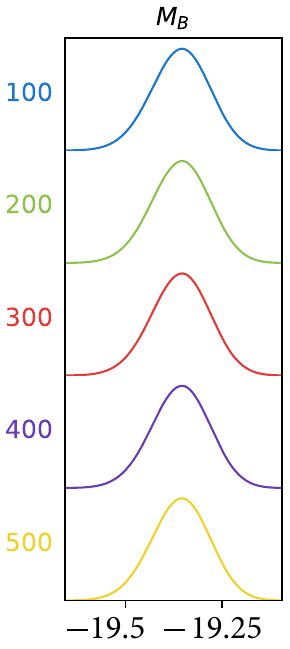}
\caption{Marginalized posteriors of supernovae apparent magnitude $M_B$ as a function of the number of considered ANN reconstructions. The obtained constraints obtained are $M_B$ = $-19.352^{+0.073}_{-0.079}$, $-19.352^{+0.073}_{-0.079}$, $-19.353^{+0.073}_{-0.078}$, $-19.353^{+0.073}_{-0.078}$, and $-19.353^{+0.073}_{-0.078}$, corresponding to 100, 200, 300, 400, and 500 ANN reconstructions.} \label{fig:MB_compare}
\end{figure}

\subsection{Constraints on \texorpdfstring{$M_B$}{} \label{subsec:constraints}}

With our trained network model, we proceed to reconstruct $m$ and its 1$\sigma$ uncertainty $\Delta m$ at the CC redshifts. We also obtain the derivatives of $m$ at the respective redshifts, hereafter denoted as $m'$, directly employing the automatic differentiation module (namely \texttt{torch.autograd.grad}) to the distribution of predictions obtained during inference. This approach thus helps to simultaneously compute the mean and 1$\sigma$ uncertainty of the reconstructed functions. The plots for the reconstructed $m(z_{\text{CC}})$ and $m'(z_{\text{CC}})$ is shown in Fig. \ref{fig:mz_cc_recon}. Here we also show the distribution of $n$-predicted $m(z)$ and $m^\prime(z)$ ANN reconstruction samples. For illustration, we have set $n$=100. The normalized covariance matrices between the reconstructed functions can be visualized from Fig. \ref{fig:mz_cc_cov_recon}. 

For these $n$-reconstructed $m$ and $m'$, one can write down the corresponding Hubble parameter as a function of $M_B$ utilizing Eq. \eqref{eq:Hubble_from_sn}. Finally, to obtain the model-independent constraints on $M_B$ we define this $\chi^2$ function, 
\begin{equation} \label{eq:chi_MB}
    \chi^2 \left\lbrace M_B \right\rbrace = \sum_{i,j}  \left[H_{\text{CC}}(z_i) - H_{\text{ANN}}(z_i)\right]^{\text{T}} \Sigma^{-1}_{ij} \left[ H_{\text{CC}}(z_j) - H_{\text{ANN}}(z_j) \right] \, .
\end{equation}
The corresponding log-likelihood is thus given by,
\begin{equation}\label{eq:log-likelihood}
    \log \mathcal{L} \left\lbrace M_B \right\rbrace = - \frac{\chi^2}{2} - \frac{1}{2} \log \vert \Sigma \vert - \frac{N}{2} \log(2\pi) \, ,
\end{equation}
where $N = 32$, the size of the CC data. Instead of opting for a $\chi^2$ minimization, we undertake a Markov Chain Monte Carlo analysis, where we maximize the log-likelihood by minimizing the negative log-likelihood considering a uniform prior for $M_B \in [-21, -18]$.

Initially, we start with $n$=100 ANN-predictions of $m(z_{\text{CC}})$ and $m^\prime(z_{\text{CC}})$. Furthermore, we increase $n$ iteratively in steps of 100 up to $n=500$, i.e., $n$=100, 200, 300, 400, 500 respectively. We find that the resulting constraint on $M_B$ is slightly dependent on $n$, the total number of ANN reconstructions for $m(z)$ considered, as illustrated in Fig. \ref{fig:MB_compare}, where a statistically stable value of $M_B=-19.353^{+0.073}_{-0.078}$ was reached at around $n=300$ predictions.

\begin{table}
{\renewcommand{\arraystretch}{1.25} \setlength{\tabcolsep}{5 pt} \centering \scriptsize
\begin{tabular}{l c c c}
\hline          
Reference &  Methodology &  Datasets & $M_B$ \\ 
\hline 
\hline
\multirow{2}{*}{Camarena \& Marra\cite{Camarena:2019rmj}} & \multirow{2}{*}{Cosmography} &  $\alpha_{\text{BAO}}$ + $r_d^{\text{CMB}}$ + Pantheon & $-19.401 \pm 0.027$ \\ 
 &  &   $\theta_{\text{BAO}}$ + $r_d^{\text{CMB}}$ + Pantheon  & $-19.262 \pm 0.030 $ \\ 
 \hline
\textbf{Mukherjee \& Mukherjee}\cite{Mukherjee:2021kcu} & \textbf{Gaussian Process}  & \textbf{CC + Pantheon } & $\mathbf{-19.387 \pm 0.060}$ \\ 
 \hline 
\multirow{2}{*}{\textbf{Mukherjee \& Banerjee}\cite{Mukherjee:2021epjc}} &  \multirow{2}{*}{\textbf{Gaussian Process}}  & \textbf{CC + Pantheon } & $\mathbf{-19.360 \pm 0.059}$ \\ 
 &   &  CC + $r_{\text{BAO}}$ + Pantheon  & $-19.398 \pm 0.041$ \\ 
 \hline 
\multirow{3}{*}{\textbf{Dinda \& Banerjee}\cite{Dinda:2022jih}} & \multirow{3}{*}{\textbf{Gaussian Process}}  & \textbf{CC + Pantheon } & $\mathbf{-19.384 \pm 0.052}$ \\ 
 &   &  BAO + Pantheon & $-19.396 \pm 0.016$ \\ 
 &   &  CC + BAO + Pantheon & $-19.395 \pm 0.015$ \\ 
\hline   
 \multirow{2}{*}{Benisty \textit{et al.}\cite{Benisty:2022psx}}  &  \multirow{2}{*}{Neural Networks} &  $\alpha_{\text{BAO}}$ + $r_d^{\text{CMB}}$ + Pantheon  & $-19.38 \pm 0.20$ \\
  &    &  $\alpha_{\text{BAO}}$ + $r_d^{\text{SH0ES}}$ + Pantheon  & $-19.22 \pm 0.20$ \\
\hline
\multirow{2}{*}{G\'{o}mez-Valent\cite{Gomez-Valent:2021hda}} & \multirow{2}{*}{Index of Inconsistency} & CC + BAO + Pantheon & $-19.374 \pm 0.080$ \\
 &  & CC + BAO + Pantheon ($\Omega_k \neq 0$) & $-19.362^{+0.078}_{-0.067}$ \\
 \hline 
\multirow{4}{*}{Favale \textit{et al.}\cite{Favale:2023lnp}} & \multirow{4}{*}{Gaussian Process} & CC + Pantheon+  & $-19.344^{+0.116}_{-0.090}$ \\
 &  &   CC + BAO + Pantheon+   & $-19.314^{+0.086}_{-0.108}$ \\ 
 &  &   CC + SH0ES + Pantheon+  & $-19.252^{+0.024}_{-0.036}$ \\
 &  &   CC + BAO + SH0ES + Pantheon+ & $-19.252^{+0.024}_{-0.036}$ \\
\hline   
Banerjee \textit{et al.}\cite{Banerjee:2023evd} & Gaussian Process & CC + $r_{\text{BAO}}$ + Pantheon+ ($\Omega_k \neq 0$) & $-19.404^{+0.099}_{-0.104}$ \\
\hline 
\multirow{2}{*}{Shah \textit{et al.}\cite{Shah:2024slr}} & \multirow{2}{*}{Neural Networks} & Pantheon  + $\alpha_{\text{BAO}}$ + $r_d^{\text{CMB}}$ & $-19.394^{+0.018}_{-0.017}$ \\
 &  & Pantheon  + $\theta_{\text{BAO}}$ + $r_d^{\text{CMB}}$ & ${-19.257}^{+0.028}_{-0.027}$ \\
\hline
\textbf{This work$^{\star}$} &  \textbf{Neural Networks} & \textbf{CC \& Pantheon+ } & $\mathbf{-19.353^{+0.073}_{-0.078}}$ \\ 
\hline         
\end{tabular}
\caption{Comparison between the model-independent constraints on $M_B$ obtained in this work vs those present in the literature.}
\label{tab:comparison_MB}
}
\end{table}

\begin{figure}
\centering
\includegraphics[width=0.75\textwidth]{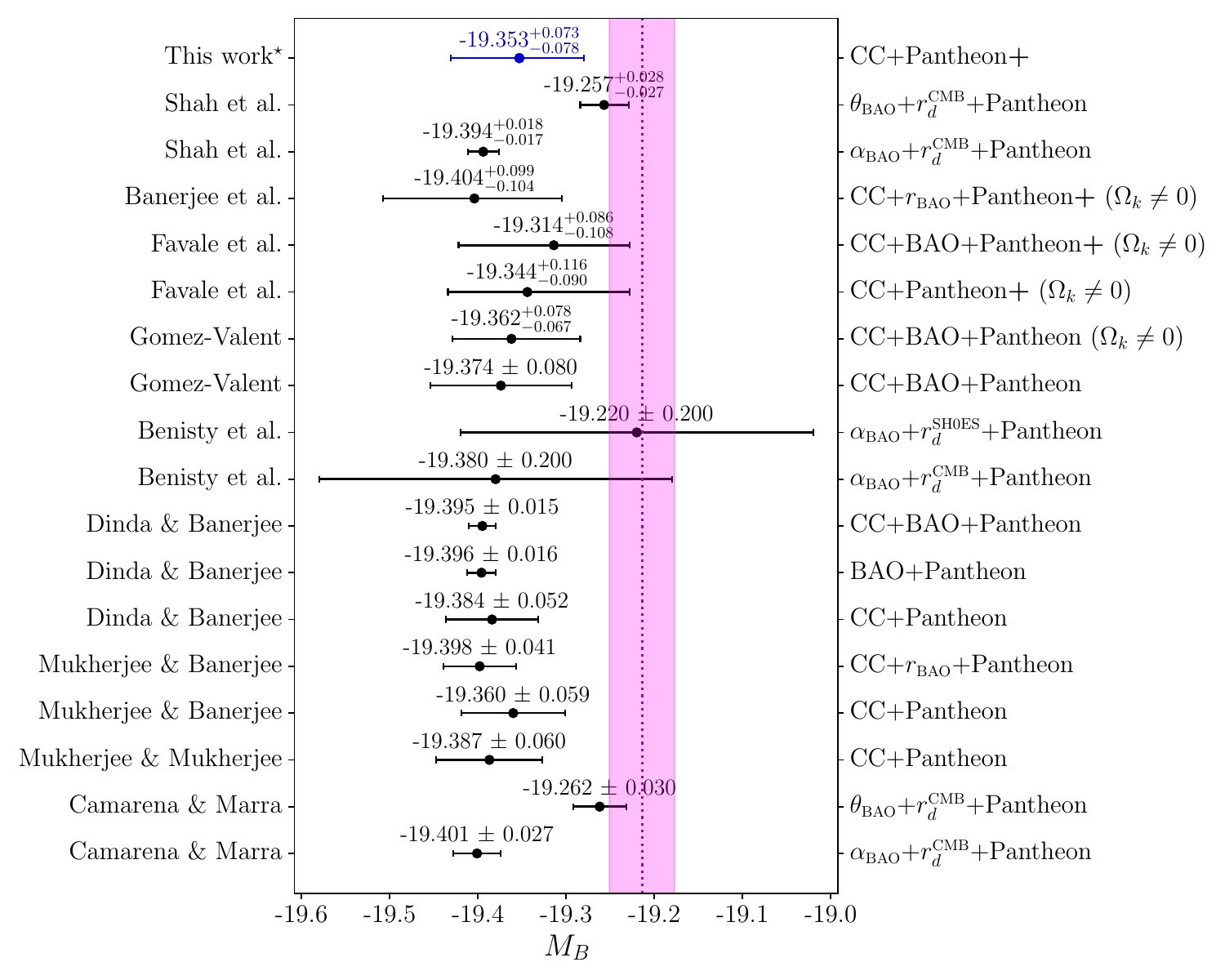}
\caption{Whisker plot showing a comparison between the model-independent constraints on $M_B$ obtained in this work vs those present in the existing literature. The purple region corresponds to the $1-\sigma$ model independent constraint $M_B = -19.214 \pm 0.037$, as reported in Ref. \cite{Riess:2020fzl, Riess:2021jrx} inferred from the Pantheon and SH0ES data sets. }
\label{fig:MB_whisker}
\end{figure}

There have been similar efforts to compute $M_B$ from different combinations of cosmological observations \cite{Camarena:2019rmj, Dinda:2022jih, Banerjee:2023evd, Favale:2023lnp, Mukherjee:2021epjc, Shah:2024slr, Gomez-Valent:2021hda, Mukherjee:2021kcu}. For a comparative analysis, we mention some of the previous results, for a given set of observational data, in Table \ref{tab:comparison_MB}. Finally, a whisker plot is shown with the available results in Fig. \ref{fig:MB_whisker} for better illustration. We find that our results are consistent with the constraints on $M_B$ obtained from almost all the previous analyses, particularly when similar data sets are taken into consideration \cite{Mukherjee:2021epjc, Dinda:2022jih, Mukherjee:2021kcu}. It deserves to be mentioned that the constraints on $M_B$ are tighter when the BAO observations are combined with the CC and SNIa data. This is because the uncertainty associated with the BAO data set is significantly smaller in comparison to the CC data set. However, in this work, we refrain from including the BAO data sets because there are apprehensions regarding their dependence towards some fiducial cosmological model. So, we undertake this analysis utilizing only the SNIa and CC data sets to ensure a complete model-independent prescription.

\begin{table}
{\renewcommand{\arraystretch}{1.3} \setlength{\tabcolsep}{20 pt} \centering \small
    \begin{tabular}{c c}
        \hline
		$\bar{z}$ & $M_B(\bar{z})$ \\ 
		\hline
		0.275 & $-19.404^{+0.097}_{-0.104}$ \\ 
		0.746 & $-19.61^{+0.15}_{-0.16}$ \\ 
		1.234 & $-19.14^{+0.14}_{-0.15}$ \\ 
		1.748 & $-19.46\pm 0.22$ \\ 
		\hline
    \end{tabular}
    \caption{Inferred constraints on $M_B(\bar{z})$ using the redshift layers binning technique.}
    \label{tab:constraints_table}
}
\end{table}

\begin{figure}
\centering
\includegraphics[width=0.485\textwidth]{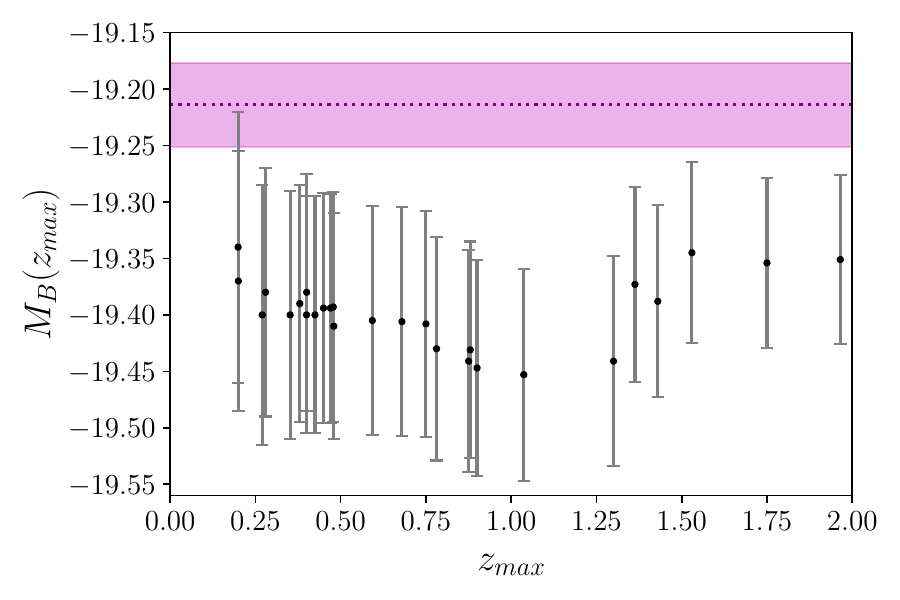}
\includegraphics[width=0.485\textwidth]{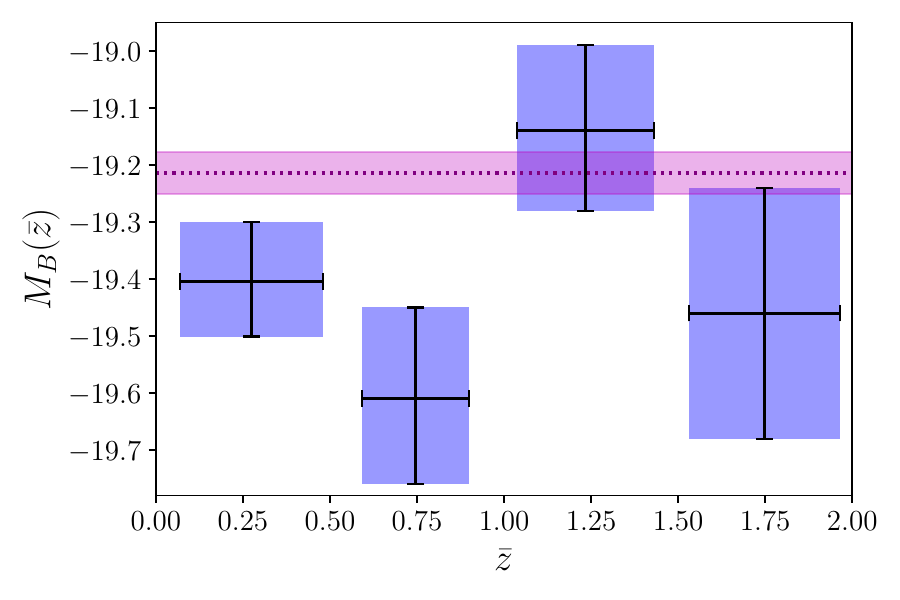}
\caption{Predictions of the supernovae absolute magnitudes: $M_B(z_\mathrm{max})$ by adopting cumulative binning, where $M_B(z_\mathrm{max})$ is the derived value of $M_B$ by considering CC $H(z)$ data up to $z_\mathrm{max}$ (left panel), and $M_B(\bar{z})$ by adopting the redshift layer binning, where $M_B(\bar{z})$ is the derived value of $M_B$ by considering CC $H(z)$ data within that redshift layer with a mean redshift $\bar{z}$ (right panel). The purple region corresponds to the $1-\sigma$ model independent constraint $M_B = -19.214 \pm 0.037$, as reported in Ref. \cite{Riess:2020fzl, Riess:2021jrx} inferred from the Pantheon and SH0ES data sets.} \label{fig:MB_cumulative}
\end{figure}

\subsection{Data-driven transition of \texorpdfstring{$M_B$}{} \label{subsec:transition}} 

We now investigate the possibility of a redshift evolution in $M_B$ by considering two types of binning methods, applied to the adopted $H(z)$ CC data which was already used in Sec. \ref{subsec:constraints}. The first binning method is the so-called cumulative binning technique \cite{Hu:2022kes,Jia:2022ycc}, in which the next redshift bin has one more high-redshift $H(z)$ data point with respect to the previous redshift bin. To be able to run the MCMC analysis, in our first cumulative redshift bin, we considered the first six low-redshift $H(z)$ data points up to $z_\mathrm{max}=0.1993$. The derived values of $M_B(z_\mathrm{max})$ from the cumulative binning method are shown in Fig. \ref{fig:MB_cumulative}, where $M_B(z_\mathrm{max})$ denotes the value of $M_B$ inferred from the $H(z)$ data set with maximal redshift $z_\mathrm{max}$.

The other binning strategy that is considered in this work is the approach of conducting the MCMC analysis to determine $M_B$ by considering redshift layers \cite{Pasten:2023rpc}, where each layer has a mean redshift $\bar{z}$. In our analyses, we have considered four layers with $\bar{z}=[0.275,\,0.746,\,1.234,\,1.748]$, and within each redshift layer we have inferred a redshift-dependent constraint on $M_B$ as outlined in Table \ref{tab:constraints_table}. 

From the outcomes of the redshift-dependent constraints of $M_B$, it should be remarked that there is an indication of a transition in the value of $M_B$ at $z\approx1$, irrespective of the binning techniques adopted.

\section{Conclusion \label{sec:conclusion}}

In the standard practice, when estimating the cosmic distances as a function of the scale factor and its derivatives, the intrinsic luminosity of Type Ia supernovae is considered a standard candle. This assumption is of utmost importance in understanding the accelerated expansion of the Universe. In this study, we have tried to investigate if this assumption is as educated as it sounds, by constraining the peak absolute magnitude, $M_B$, of SNIa, in a cosmological model-independent way. For this exercise, we have utilized two distinct sets of observational data, namely the Pantheon+ SNIa compilation and the cosmic chronometer Hubble parameter measurement, as described in Sec. \ref{subsec:datasets}.

In this work, we have explored the nature of the absolute magnitude parameter through the use of ANNs with a focus on obtaining a model-independent value of this parameter, as well as assessing whether this parameter exhibits a statistically relevant evolution. We demonstrate how the Pantheon+ sample can be utilized, through Eq.~\ref{eq:Hubble_from_sn}, to determine the Hubble parameter at different redshift points. This information is then used in conjunction with the CC Hubble parameter observations to constrain the value of $M_B$ in a model-independent manner, employing the log-likelihood defined in Eq.~\ref{eq:log-likelihood}. Our results, summarized in Table \ref{tab:comparison_MB}, indicate that the final constraint on $M_B$ is competitive with values in the existing literature. Additionally, this approach had the advantage of not relying on a fiducial cosmological model. 

Since the focus is on constraining $M_B$ in a model-agnostic way, we do not consider any parametric or theoretical form between $d_L$ and $H(z)$ at the outset. Instead, given the observed apparent magnitudes $m$ of the SNIa, we proceed to reconstruct $m(z)$ a function of the redshift employing Artificial Neural Networks (Sec. \ref{subsec:training}). Utilizing the equations in Sec. \ref{subsec:theory}, the analogous Hubble parameter, inferred from SN-Ia via Eq.~\eqref{eq:Hubble_from_sn}, can in principle be written as a function of its peak absolute magnitude $M_B$. Finally, having defined the $\chi^2$ function for $M_B$ (Eq. \eqref{eq:chi_MB}), we obtain constraints by minimizing the negative log-likelihood in Eq.~\eqref{eq:log-likelihood}. The result we get is 
\begin{equation}
M_B=-19.353^{+0.073}_{-0.078} \, .
\end{equation}

Furthermore, we proceed to test the evolution of the peak absolute magnitude $M_B$ as a function of redshift. For this exercise, two types of binning methods have been adopted while working with the CC data. In the first one, we used the cumulative method, in which a redshift bin has one more data point with respect to the previous bin and the results are shown in Fig.~\ref{fig:MB_cumulative}. In the second one, we considered four distinct binning of redshifts, where each one has a mean redshift $\bar{z}$, and the results are presented in Table~\ref{tab:constraints_table}. In both cases, one can notice a transition in the value of $M_B$ at $z \approx 1$. Given that both methods result in consistent transitional redshifts, this indicates that the result is likely to be a genuine result of the underlying pipeline method rather than a statistical irregularity. To further validate our reconstruction approach and analysis methodology, we have added an appendix \ref{sec:mock} dedicated to a thorough test of our statistical pipeline. This includes further mock simulations specifically crafted to assess the estimation of \( M_B \) and to detect potential redshift variations in \( M_B \) as a function of \( z \). We considered two different scenarios for injecting \( M_B \) to create the Pantheon+-like mock apparent magnitude $m(z)$ data. This extensive testing enhances the confidence in the robustness and reliability of our method. 

More broadly, previous works in the literature suggest the possibility of a transition redshift at which certain parameters, including the $M_B$ parameter, may suffer a change in value. Ref.~\cite{Akarsu:2023mfb,Paraskevas:2024ytz} points to a sign switch in the cosmological constant from positive to negative, while Ref.~\cite{Perivolaropoulos:2021bds, Lovick:2023tnv} indicates a transition or evolution in the value of $M_B$ albeit at a much closer distance. Indeed, even a regular MCMC albeit with a binning procedure will produce a variety of $H_0$ values as shown in Refs.~\cite{Colgain:2023bge,Colgain:2022tql,Dainotti:2021pqg,Dainotti:2022bzg,Dainotti:2023bwq,Hu:2023jqc} some of which may ultimately be interpreted as a variation in the $M_B$ parameter.

There are indications towards a variation in the value of $M_B$ present in the existing literature \cite{Efstathiou:2021ocp,Camarena:2022iae,Camarena:2021jlr}, which may represent the underlying physical nature of the tension in terms of astrophysics. Understanding the source and characterizing the $H_0$ tension may express other important features relevant for further study. In future work, we hope to apply novel machine learning approaches to assessing the evolution profile of other important, rather critical, parameters that significantly contribute to or influence the constraints on the value of the Hubble constant.

\appendix

\section{Method Validation with Mock Simulation \label{sec:mock}}

\begin{figure}
\centering
\includegraphics[width=0.45\textwidth]{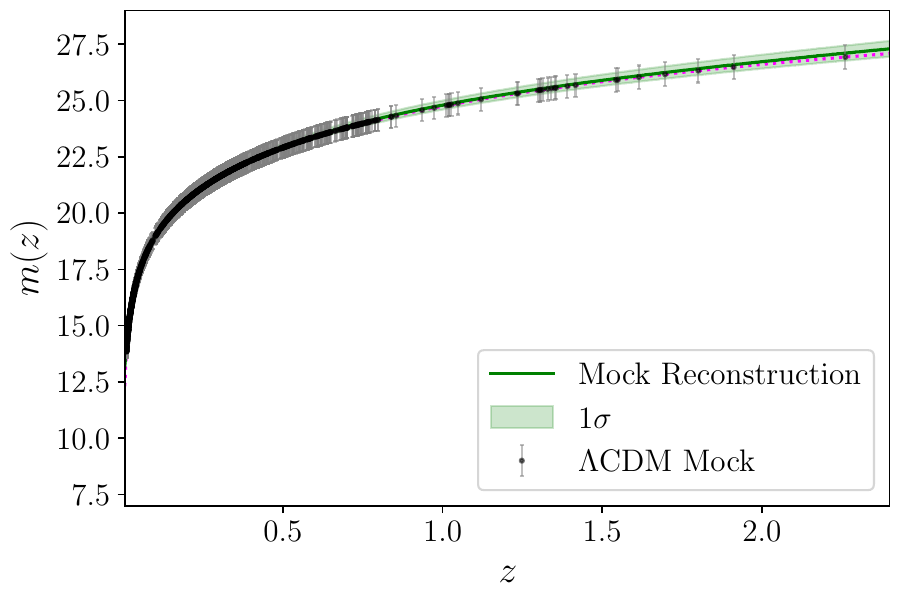} 
\includegraphics[width=0.45\textwidth]{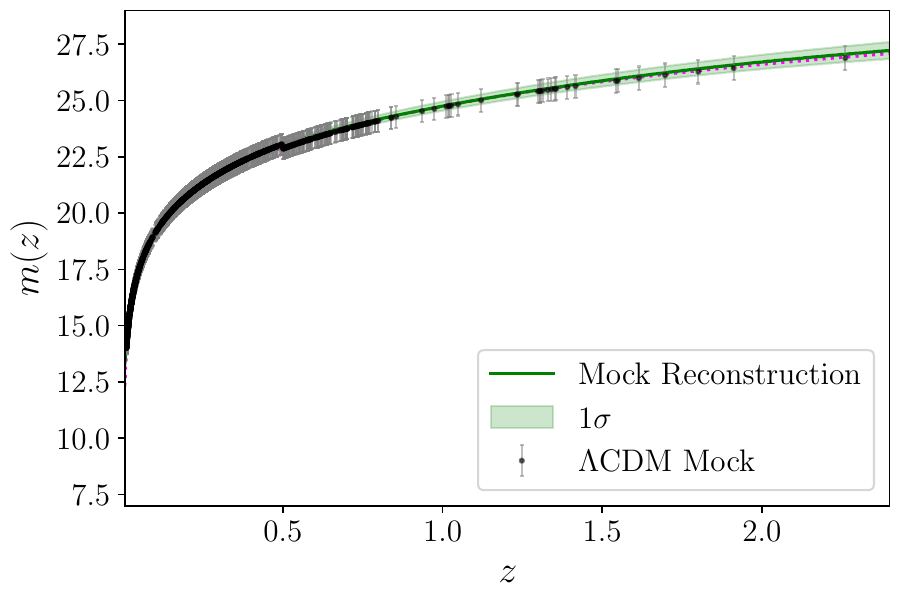} 
\caption{Reconstruction of the simulated mock Pantheon+ like SNIa apparent magnitudes $m(z)$: for Case (i) a constant $M_B = -19.4$ for all redshifts $z$ [in left panel] and Case (ii) with a transition viz. $M_B = -19.4$ for $z \geq 0.5$ and $M_B = -19.2$ for $z < 0.5$. [in right panel].} \label{fig:mz_mock_plot}
\end{figure}

To validate our pipeline, we use a fiducial $\Lambda$CDM cosmology with parameters $\Omega_{m0} = 0.3$ and $H_0 = 70 \, \text{km} \, \text{Mpc}^{-1} \text{s}^{-1}$. We generate mock data using the Pantheon+ compilation and CC Hubble parameter measurements. For the mock Pantheon+ data, we perform two types of injections for the absolute magnitude value: (i) a constant $M_B = -19.4$ for all redshifts $z$, and (ii) a transition where $M_B = -19.4$ for $z \geq 0.5$ and $M_B = -19.2$ for $z < 0.5$. To account for uncertainties, we introduce random Gaussian noise corresponding to 2\% of the magnitude values for both mock $m(z)$ and $H(z)$ respectively.

Initially, we train our network model using this mock Pantheon+ samples. The results of this training are depicted in Fig. \ref{fig:mz_mock_plot}, following the procedure outlined in Sec. \ref{subsec:training}. Our primary objectives are twofold: First, we aim to infer the value of \( M_B \) used to create the mock Pantheon+-like catalogue in conjunction with the mock CC $H(z)$ data, as discussed in Sec. \ref{subsec:constraints}. We find that the resulting constraints on the absolute magnitude for Case (i) is $M_B = -19.39^{+0.053}_{-0.052}$, which includes the fiducial injected value of -19.4 within the 1$\sigma$ confidence level. For Case (ii), which assumes a simulated mock transition in the absolute magnitude, we obtain $M_B = -19.31^{+0.045}_{-0.046}$.

\begin{figure}
\centering
\includegraphics[width=0.45\textwidth]{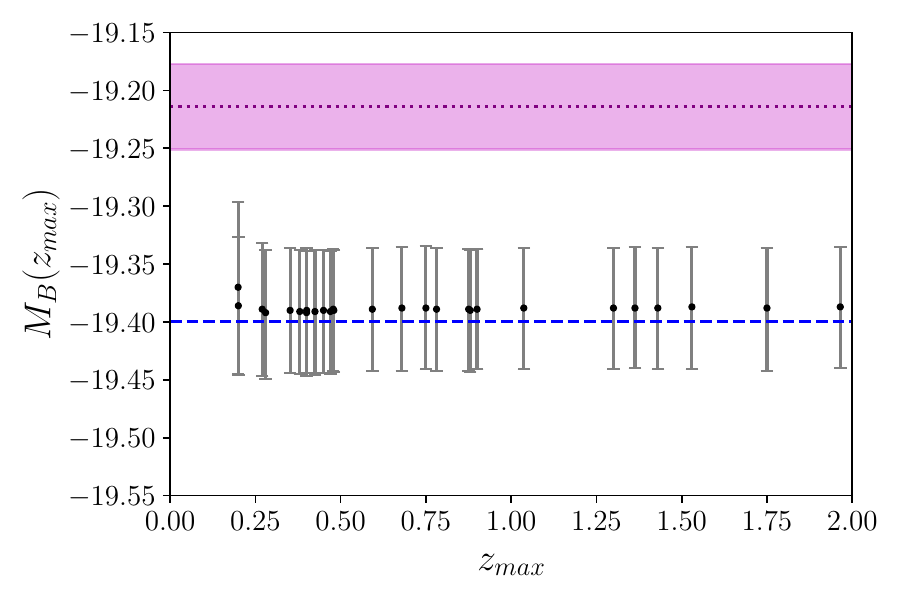}
\includegraphics[width=0.45\textwidth]{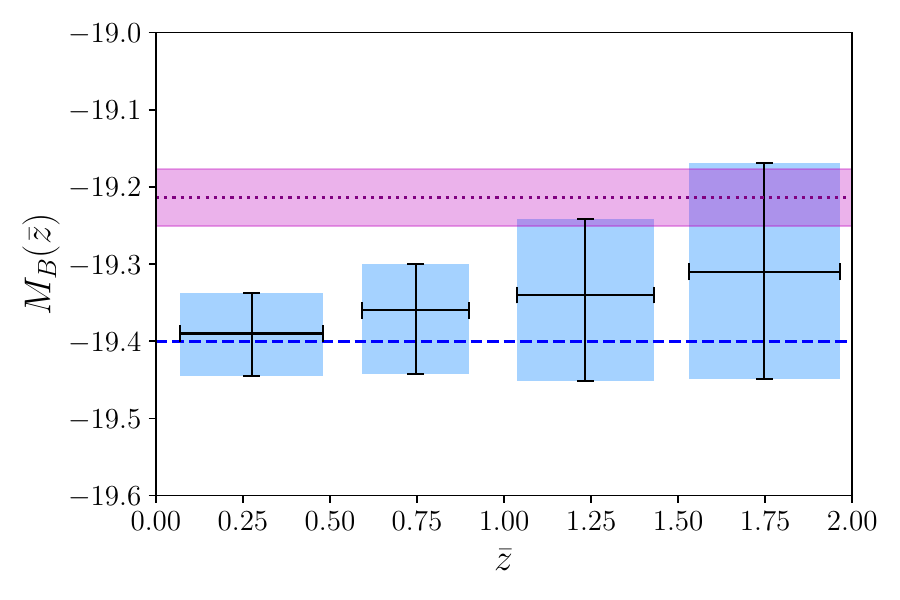} \\
\includegraphics[width=0.45\textwidth]{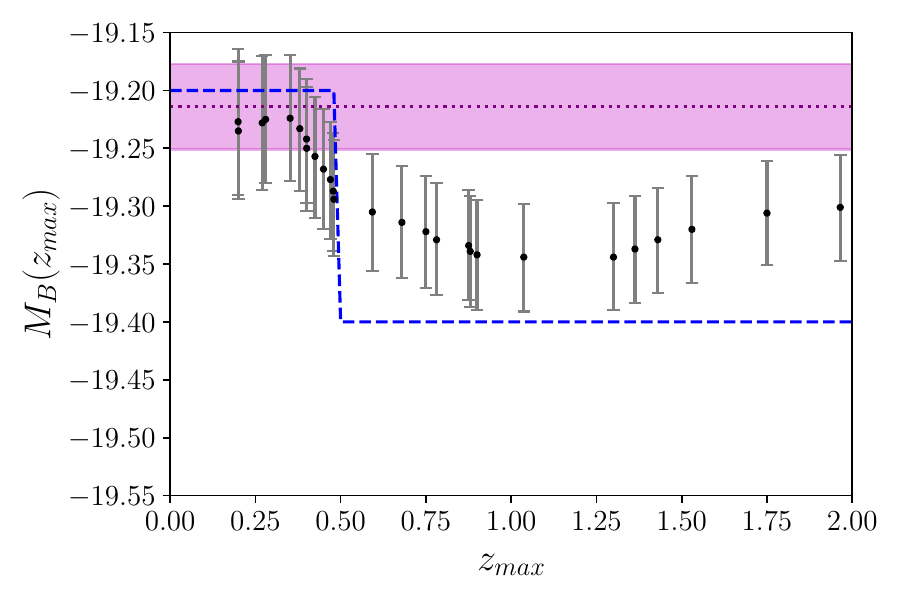}
\includegraphics[width=0.45\textwidth]{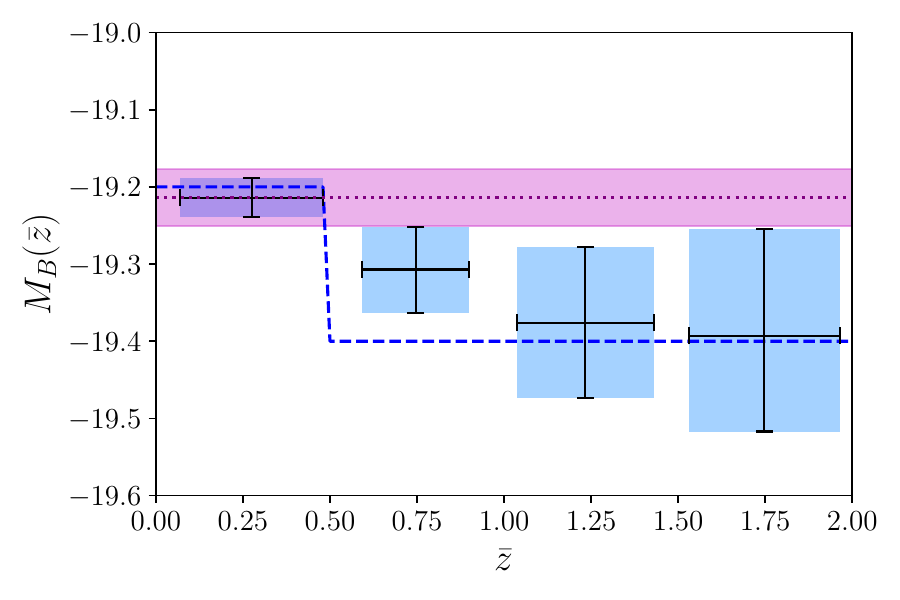}
\caption{Predictions for the mock SNIa absolute magnitudes: $M_B(z_\mathrm{max})$ by adopting cumulative binning (left panel), and $M_B(\bar{z})$ by adopting the redshift layer binning (right panel). The top panel shows Case (i) injection of a constant $M_B = -19.4$ for all $z$, whereas the bottom panel shows Case (ii) injection with a transition viz. $M_B = -19.4$ for $z \geq 0.5$ and $M_B = -19.2$ for $z < 0.5$. The blue dashed line shows the injected $M_B$ values. The purple region corresponds to $M_B = -19.214 \pm 0.037$ \cite{Riess:2020fzl, Riess:2021jrx}.} \label{fig:test_mock_plot}
\end{figure}

Second, we aim to investigate the potential redshift evolution of \( M_B \) through a null hypothesis test, utilizing both cumulative and layered binning methods as outlined in Sec. \ref{subsec:transition}. In Fig. \ref{fig:test_mock_plot}, we present the predictions of simulated SNIa absolute magnitudes for both Case (i) and Case (ii) mock $M_B$ injections. The left panel displays the predictions of $M_B(z_\mathrm{max})$ using cumulative binning, where $M_B(z_\mathrm{max})$ is the value derived from mock $H(z)$ data up to $z_\mathrm{max}$. The right panel shows the constraints on $M_B(\bar{z})$ obtained through redshift layer binning, where $M_B(\bar{z})$ is derived from mock $H(z)$ data within a redshift layer centred around a mean redshift $\bar{z}$.

The top panel shows that the predicted values of \( M_B \) remain constant, with the injected value of -19.4 consistently within the 1$\sigma$ confidence level. In contrast, the bottom panel displays variations in the predicted values of \( M_B \) across different redshifts, regardless of the binning strategy. This variation aligns with the injected values of \( M_B \): -19.4 for \( z \geq 0.5 \) and -19.2 for \( z < 0.5 \). Although the variation is smoother rather than the sharp transition that was injected, our findings suggest that the pipeline can effectively be used to test for redshift-dependent changes in \( M_B \).

\begin{acknowledgments}

This paper is based upon work from COST Action CA21136 {\it Addressing observational tensions in cosmology with systematics and fundamental physics} (CosmoVerse) supported by COST (European Cooperation in Science and Technology). We thank the anonymous reviewer for their valuable suggestions towards the improvement of the manuscript. PM acknowledges the financial support from ISI Kolkata and DST SERB, Government of India under the National Post-Doctoral Fellowship (File No. PDF/2023/001986). JLS and JM would also like to acknowledge funding from ``The Malta Council for Science and Technology'' as part of the ``FUSION R\&I: Research Excellence Programme'' REP-2023-019 (CosmoLearn) Project. The work was supported by the PNRR-III-C9-2022–I9 call, with project number 760016/27.01.2023. 

\end{acknowledgments}

%%%%%%%%%%%%%%%%%%%%%%%%%%%%%%%%%%%%%%%%%%%%%%%%
%%%%%%%%%%%%%%%%%%%%%%%%%%%%%%%%%%%%%%%%%%%%%%%%

\bibliographystyle{JHEP}
\bibliography{references.bib}

\end{document}